**Aging and Communication in the Twin Paradox**

David A. de Wolf[*]
*W. Halfrond, 431, 1183 JD Amstelveen, the Netherlands*

Abstract

The twin paradox of the special theory of relativity has given rise to a large body of literature discussing its implications. In its standard form, the traveler changes velocity only at the destination of the trip, so that he appears to perceive an improbably instantaneous and non-continuous change in age of the stationary twin. In this work, a smooth velocity/acceleration profile is used that allows the abrupt velocity-change case as a limit. All gravitational effects are ignored in this treatment. Aside from mutual perception of simultaneous clock times in an accelerating frame, constant communication of clock times between the twins by means of (digital) light signals is shown to be possible, in principle if not in practice.

## I. INTRODUCTION

In the standard twin paradox[1,2,3,4,5,6,7] of the special theory of relativity[8], a space traveler 𝔅 travels at constant velocity $v (= \beta c)$ from a point on Earth to a distant star and returns with the same velocity dependence to Earth. This round trip gives rise to a remarkable relativistic phenomenon: 𝔄 ages less than 𝔅 (as perceived by 𝔅) on each leg of the trip, yet at the end of the entire trip 𝔄 has aged more than 𝔅! A discontinuously rapid aging of 𝔄 must therefore occur instantaneously at the turn-around point. There is some experimental evidence[9,10] for the discrepancy in age, and therefore little argument in the literature about its correctness (although there is also some criticism[11,12]). The perception by the traveler of a discontinuous change in the Earth-bound twin's age is reasonably well understood[13,14]; it is due to a sudden change in simultaneity of events for 𝔄 and 𝔅 because 𝔅 enters a new and different inertial frame, but it is still somewhat astonishing. It is therefore illuminating to study a simple velocity profile that allows the velocity to change direction smoothly and continuously with the ability to include the discontinuous case in a parameter limit. Various other versions of the twin paradox have been considered[15,16,17,18]. These, however, do not adequately address the issues that follow.

---

[*]Emeritus professor of electrical engineering, Virginia Tech, Blacksburg, VA (*dadewolf@vt.edu*)

A separate issue in this work is communication between the twins with light signals so that each can verify his/her perception of the other's clock times. For these reasons the following variant is proposed:

## II. AN ALTERNATIVE VELOCITY PROFILE

Instead of traveling at constant absolute velocity $v$, an alternative analytical velocity profile is suggested in which adjustable parameters can vary the length of the trip, the maximum velocity, and the acceleration at the turn-around point of the trip. Unfortunately, it does not seem possible to invent a smooth and continuous one with compact support (so that the trip always starts and ends with zero velocity) that includes the standard constant-velocity case. This would be possible with a purely numerical profile, but then manipulation of the three parameters is cumbersome. Hence the following velocity profile for traveler $\mathfrak{B}$ is proposed:

$$\tfrac{v(t)}{c} \equiv \beta(t) = \tfrac{1}{2}\beta_0\{\tanh[b(t-t_s)]-2\tanh[b(t-2t_s)]+\tanh[b(t-3t_s)]\} \tag{1a}$$

While the path remains a straight line between Earth and the star, variation of parameter $t_s$ determines the length of the trip and variation of acceleration parameter $b$ allows for a range of profiles with a continuous and smooth change from $+v_{\max}$ to $-v_{\max}$ (with $v_{\max}/c \leq \beta_0$) to an abrupt change in the limit $b \to \infty$.

However, zero velocity is attained only at $t = \pm \infty$ which would render one-way travel time $T$ to be infinite. Instead, $T$ is chosen to be the time between the maximum and minimum values of $\beta(t)$. The initial and final times are then given by

$$t_i = 2t_s - T, \quad t_f = 2t_s + T \tag{1b}$$

Several choices are illustrated in Fig. 1 (the $b = 10$ profile is indistinguishable here from the discontinuous case in which the velocity is reversed instantaneously at $t/t_s = 2$). All times are shown as ratios $t/t_s$. The initial and terminal velocities $v(t_i)$ and $v(t_f)$ are nonzero but relatively small. Figure 1 shows three such velocity profiles:

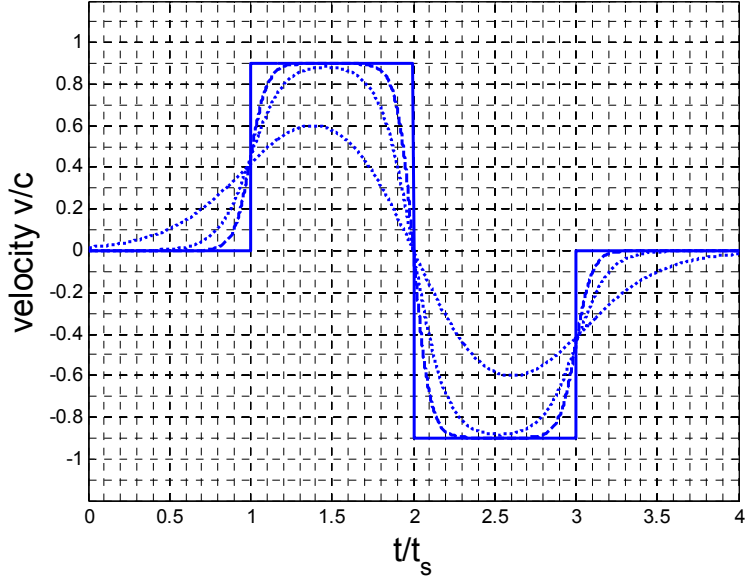

Fig. 1: Velocity profiles at $\beta_0 = 0.9$ for $b = 0.002\,(....), 0.01\,(----)$, and $10\,$ (solid);

$T(b) = 1.226\,t_s\,(....),\ 1.035\,t_s\,(----),\ t_s$ (solid),

$t_i(b) = 0.774 t_s\,(....),\ 0.965\,t_s\,(----),\ t_s$ (solid),

$t_f(b) = 3.226 t_s\,(....),\ 3.035\,t_s\,(----),\ 3t_s$ (solid).

The focus of this work is on the perception and comparison of clock times by each twin. That a clock time of one twin is perceived differently (e.g. as given by proper time for 𝔅's clock within the inertial frame of 𝔄) is, of course, a well-known consequence of the special theory of relativity, which is covered (if not always exhaustively) by all the works referred to in this article. As mentioned above, a clock time of 𝔄, as observed by 𝔅, seems to change rapidly (and even discontinuously in the limit $b \to \infty$). However, the word 'observe' needs to be considered carefully because, without extra measures, neither 𝔅 nor 𝔄 can compare clocks unless both are at one spatial location, which occurs only at $t = 0$ and $t = 2T$ of 𝔄's clock. Corresponding times on 𝔅's clock would be $t' = 0$ and $t' = 2T'$.

To overcome this problem, a second issue of this work is to include the possibility of receiving light signals (with digital clock information) from the other twin. The method followed here parallels Lasky[19] in that the monitoring occurs continuously, in this work for the velocity profiles of Fig. 1.

### III. 𝔅's clock times as observed by 𝔄

The easiest case to treat is 𝔄's calculation of 𝔅's clock time $t'$ because $t' \equiv \tau$ is the proper time for 𝔄 given, even for accelerating systems[2,3,4], by

$$t' = \int_0^t dt_1 \sqrt{(1-\beta^2(t_1))} \qquad (2)$$

which implies a continuous dependence of $t'$ upon $t$. Numerical integration of this equation for the profiles of Fig. (1) yields the curves in Fig. 2:

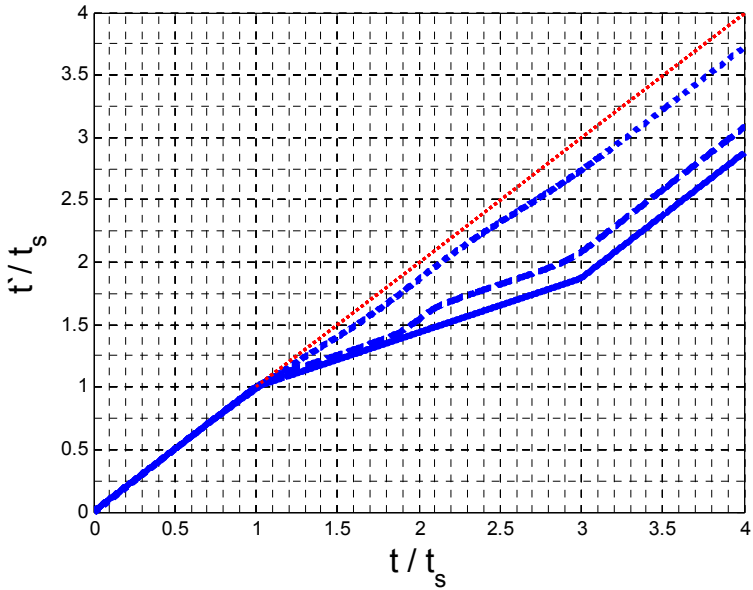

Fig. 2: 𝔅's time $t'(t)$, as deduced by 𝔄 via Eq. (2), at $\beta_0 = 0.9$ for $b = 0.002$ (....), $0.01$ (----), and 10 (solid). A $t' = t$ curve is shown for comparison. Velocity reversal occurs at $t = 2t_s$. See caption of Fig. 1 for $T$, $t_i$, $t_f$.

Similar curves for other values of $\beta$ and $b$ are easily produced but are qualitatively not different from those in Fig. 2. The time lag of 𝔅 with respect to that of 𝔄 builds up gradually and is more pronounced for profiles with higher accelerations. Furthermore, the $b \to \infty$ curve is a straight line between $t' = t_i = 1000$ and $t' = t_f = 3000$ because $\sqrt{(1-\beta^2(t'))}$ is constant.

**IV. 𝔄's clock times as observed by 𝔅**

The situation from the point of view of traveler 𝔅 is somewhat more complicated. While 𝔄 is entirely in an inertial frame, 𝔅 undergoes acceleration and is therefore not in an inertial frame. A trajectory diagram illustrates this (Fig. 3). The calculation of 𝔄's time as perceived by 𝔅 is therefore more complicated than via an equation similar to (2). Nevertheless it can be done for a small (infinitesimal) increment in time $dt'$ in which the connection to $dt$ changes form as one progresses from one to the next small increment $dt'$ in 𝔅's time and 𝔅's system is, temporarily, inertial.

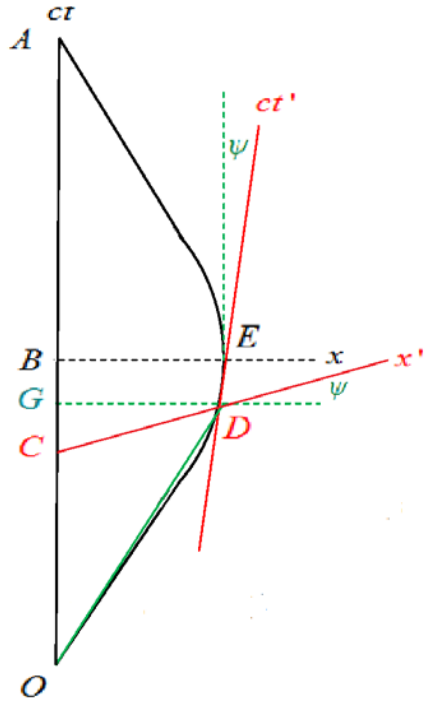

Fig. 3: Travel profiles for 𝔄 ($OGA$) and 𝔅 ($ODA$). 𝔄's Minkowski reference frame is $(x, ct)$ and 𝔅's is $(x', ct')$. The latter is shown only for point $D$.

Consider, in 𝔄's frame a Minkowski set of primed coordinate axes for each twin at point $D$ of 𝔅's trajectory, where $\psi$ is an angle to show the local tilt of the primed Minkowski coordinates at $D$. Minkowski geometry[20] dictates that $\tan \psi = \beta$. Points $D$ and $C$ are simultaneous in 𝔅's frame, suggesting via the usual Minkowski calculus, in which $\beta(t') = \tanh \psi$ instead of $\beta(t') = \tan \psi$ is used for geometrical ratios, somewhat counter-intuitively,[19] that interval $dt$ at $C$ is related to 𝔅's interval $dt'$ at $D$ by $dt = dt' \cosh \psi = [1 - \beta^2(t')]^{-1/2} dt'$ (i.e. that $dt > dt'$). However, the step from this infinitesimal interval to time $ct = OC$ is a bit more complicated because time $ct' = OD$ is along the curved (non-inertial) trajectory of 𝔅 as a result of which the interval $dt$ is constantly shifted. Spatial distance $x' \equiv CD = c \int_0^{t'} dt_1 \beta(t_1)$ because $GC = (CD)\sinh \psi$ and $OC = OG - CG = OG - x' \sinh \psi$. Furthermore $OG$ is the time $ct$ in $\mathfrak{A}$'s frame; it is related to $ct'$ by $dt = dt'[1 - \beta^2(t')]^{-1/2}$, thus $OG = c \int_0^{t'} dt_1 [1 - \beta^2(t_1)]^{-1/2}$. The result is

$$t = \int_0^{t'} dt_1 \frac{1}{\sqrt{1-\beta^2(t_1)}} - \frac{\beta(t')}{\sqrt{1-\beta^2(t')}} \int_0^{t'} dt_1 \beta(t_1) \tag{3}$$

Upon numerical integration this results in the curves of Fig. 4:

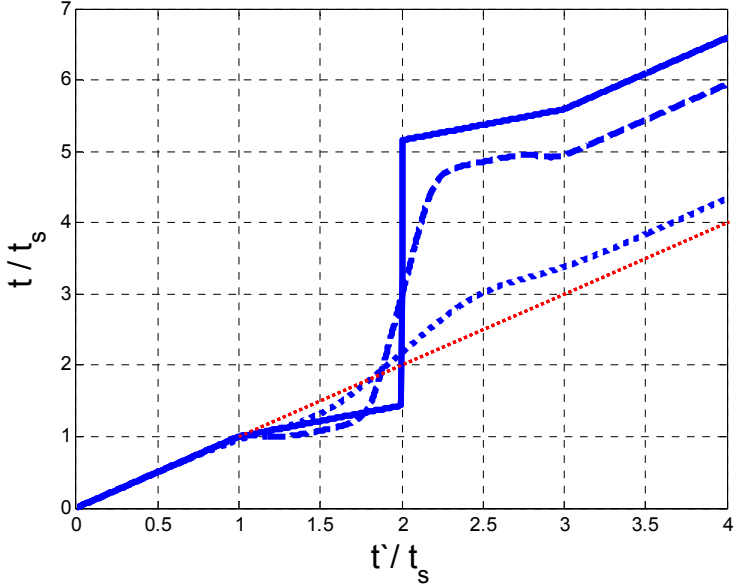

Fig. 4: $\mathfrak{A}$'s time $t(t')$, deduced by $\mathfrak{B}$ via Eq. (3), at $\beta_0 = 0.9$ for $b = 0.002$ (....), $0.01$ (----), and 10 (solid). A $t = t'$ curve is shown for comparison. Velocity reversal occurs at $t = 2t_s$. See caption of Fig. 1 for $T$, $t_i$, $t_f$.

It is not difficult to check the results for $b \to \infty$ because the integrals then are trivial.

$$\begin{array}{ll} t = t' \sqrt{1-\beta_0^2} & \text{for } t' < T' \\ t = T' \frac{1+\beta_0^2}{\sqrt{1-\beta_0^2}} + \frac{\beta_0^2}{\sqrt{1-\beta_0^2}} (T' - t') & \text{for } T' < t' < 2T' \end{array} \tag{4}$$

This is a well-known result[13], and it shows that $t$ jumps by $2\beta_0^2 T'/\sqrt{1-\beta_0^2}$ at the turn-around point $t' = T'$. The discontinuity is apparent for $b = 10$, which is indistinguishable here from the $b \to \infty$ case. More generally, the graphs show that $\mathfrak{A}$ ages more slowly than $\mathfrak{B}$ initially, at least up to $t' = T'$, and also closer to $t' = 2T'$, but $\mathfrak{A}$ ages so much more rapidly around $t' = T'$ with the result that $\mathfrak{A}$ has aged more than $\mathfrak{B}$ at $t' = 2T'$ (the time of return to Earth according to $\mathfrak{B}$).

## V. LIGHT SIGNALS FROM 𝔄 TO 𝔅

So far predictions by each twin of the other's clock time have been discussed, but these would need to be verified. This is easily done at $t = 0$, $2T$ when both twins are at the same place and can compare clocks. At intermediate times some form of communication is necessary. Most interesting here is the way that 𝔅 can monitor 𝔄's clock time continuously, and in particular by using light signals[19] sent from 𝔄 to 𝔅, as illustrated in the Minkowski diagram of Fig. 5 for $0 < t < T$.

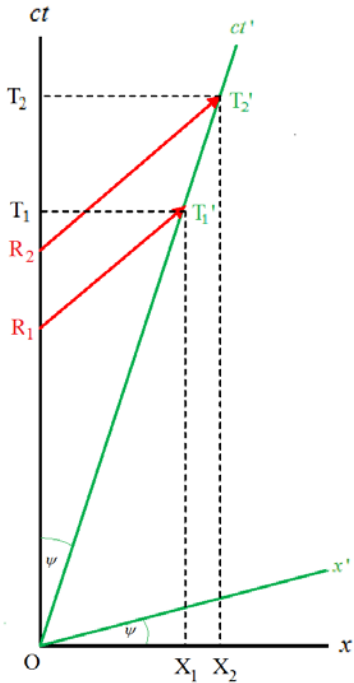

Fig. 5: Minkowski diagram for $0 < t < T$, including continuous transmission of 𝔅's clock time to 𝔄 with light signals.

Some fundamental aspects of using light signals have been discussed by Nelson[21]. Angle $\psi$ is again characterized by $\tanh \psi = \beta(t')$. Ratio definitions of angles in a Minkowski diagram can be given, as is more commonly done, by trigonometric functions but then the scales of the primed coordinates need to be adjusted by a factor $\sqrt{(1+\beta^2)/(1-\beta^2)}$. This adjustment is unnecessary when using the hyperbolic ratios. Twin 𝔅 has traveled the infinitesimal distance $X_1 X_2$ in infinitesimal time $dt' = T_1' T_2'$. A digital signal identifying time $t'$ can be sent in this way (but 𝔅 can also send a marker signal, store his measured time $t'$ and compare it to 𝔄's measured time -- see below -- upon arrival back on Earth).

From the hyperbolic geometry in a Minkowski diagram[20] it follows that the connection between $dt' = T_1'T_2'$ and $dt = R_1R_2$ is found from

$$OT_2 = (OT_2')\cosh\psi = t_2'/\sqrt{1-\beta^2}, \quad T_2R_2 = (OT_2')\sinh\psi = \beta t_2'/\sqrt{1-\beta^2} \tag{5}$$

and with $t_2 \equiv OR_2 = OT_2 - T_2R_2$ and the differentials of these quantities, one obtains

$$dt = dt'\sqrt{\frac{1-\beta(t')}{1+\beta(t')}} \tag{6}$$

Here, $dt$ is the time interval $\mathfrak{A}$ deduces of $\mathfrak{B}$'s time interval with light-signal travel time added to it (and this formula is related to that for the Doppler effect). Numerical integration then yields Fig. 6:

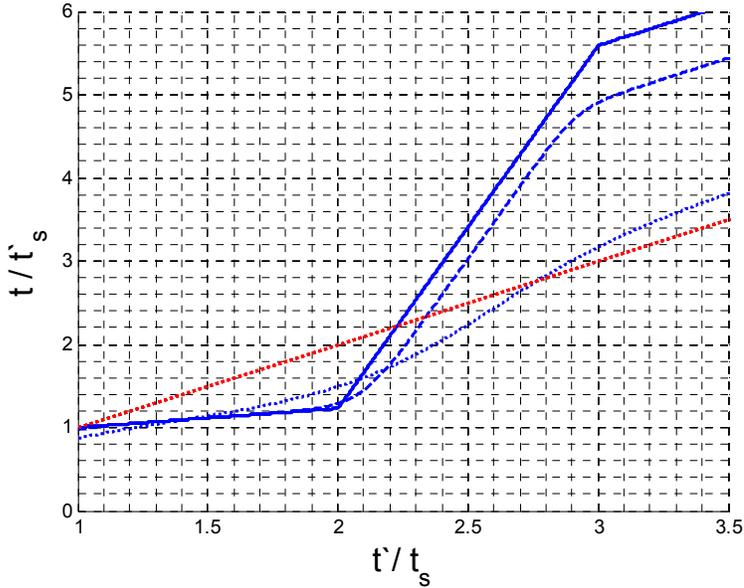

Fig. 6: $\mathfrak{A}$'s time $t(t')$, deduced by $\mathfrak{B}$ from light signals sent by $\mathfrak{A}$ to $\mathfrak{B}$ via Eq. (6), for $\beta_0 = 0.9$, $b = 0.002$ (dotted), $0.01$ (dashed), and $10$ (solid). A $t = t'$ curve is shown for comparison. Velocity reversal occurs at $t' = 2t_s$. See caption of Fig. 1 for $T$, $t_i$, $t_f$.

Figure 7 shows the same data for fixed $b = 0.01$ and various values of $\beta$:

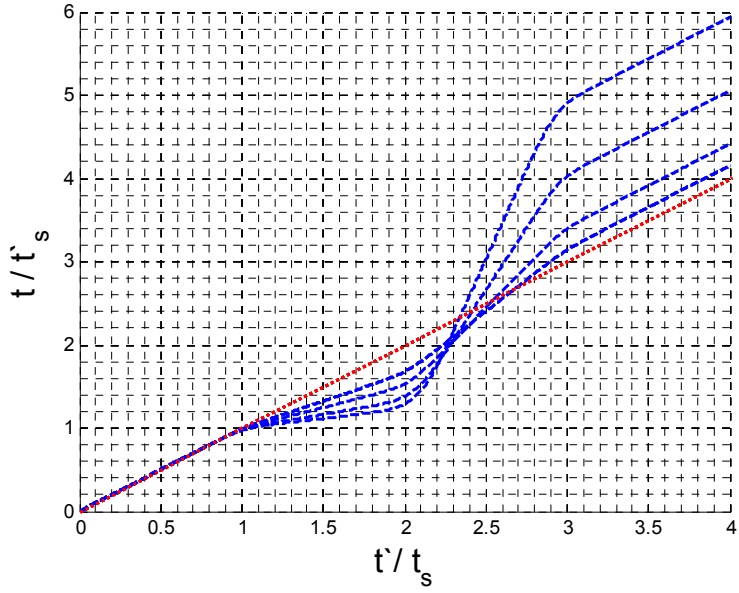

Fig. 7: 𝔄's simultaneous $t(t')$, as perceived by 𝔅, for $b = 0.01$ and various values of $\beta$, namely 0.9 (most extreme curve), 0.8, 0.6, and 0.4 (closest to $t = t'$ which is shown for comparison). Velocity reversal occurs at $t' = 2t_s$. See caption of Fig. 1 for $T$, $t_i$, $t_f$.

In this case with light signals, 𝔄 ages less rapidly than 𝔅 when $t' < T'(b)$ and more rapidly when $t' > T'(b)$, yet overall 𝔄 has aged more rapidly than 𝔅 at $t' = 2T'(b)$. More insight into this behavior is to be obtained from Fig. 8 in comparing the slopes of the curves of Fig. 6 to the $dt/dt' = 1$ line.

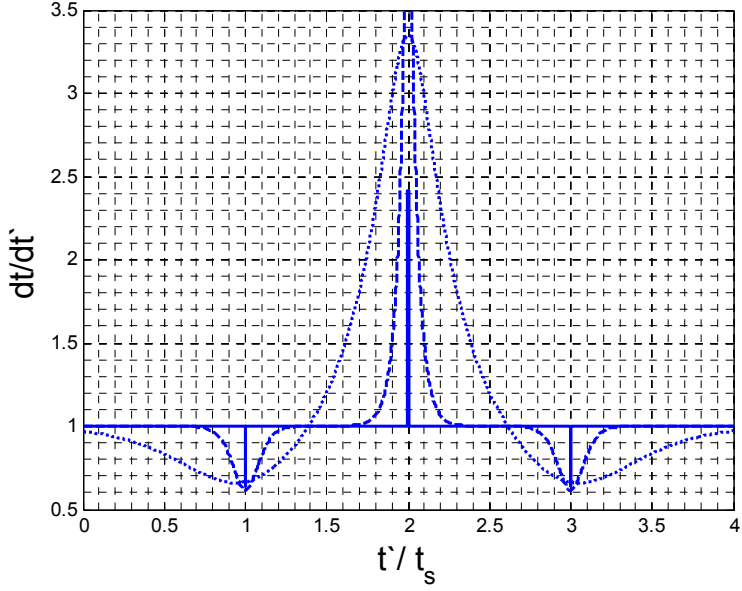

Fig. 8: Slopes of the curves in Fig. 6 (but a $b = 1$ instead of a $b = 10$ curve is shown because the latter is too narrow to be seen at present resolution). Velocity reversal occurs at $t' = 2t_s$. See caption of Fig. 1 for $T$, $t_i$, $t_f$.

Figure 8 shows at which times $dt/dt' > 1$ and when $dt/dt' < 1$. The former inequality obviously prevails and 𝔅 is perceived to have aged less than 𝔄. Results are similar for other values of $\beta$.

The contribution of the light signals yields terms of $O(\beta)$ over and above $t = t'$, whereas pure time dilation yields an effect of $O(\beta^2)$. This can be most easily seen for the $b \to \infty$ limit and for $t' < T'$ in which case the small-$\beta$ expansion yields

$$\int_0^{t'} dt_1 \sqrt{1-\beta^2} \approx t'\left[1 - \tfrac{1}{2}\beta^2 - \tfrac{1}{8}\beta^4 + O(\beta^6)\right]$$
$$\int_0^{t'} dt_1 \sqrt{(1-\beta)/(1+\beta)} \approx t'\left[1 - \beta + \tfrac{1}{2}\beta^2 - \tfrac{1}{2}\beta^3 + \tfrac{3}{8}\beta^4 + O(\beta^5)\right] \quad (7)$$

Matters are not substantially different for finite $b$ or $t' > T'$. In principle, 𝔅 can subtract the light-travel time $\beta t'$ from the second of (7) to obtain the pure time dilation as given in the first of (7). However, the light-signal effect $\beta t'$ completely swamps the time dilation (at least for $\beta \ll 1$). Hence such subtraction is probably impractical if the actual velocity profile is not known with great precision. The predominance of the light-signal effect also can be deduced from comparison of Fig. 6 with Fig. 4.

A similar analysis can be done for 𝔄's monitoring of 𝔅's clock times through light signals sent from 𝔅 to 𝔄, but this will be omitted as it is a routine analogy of the above which does not add much.

**Acknowledgements**

The author is indebted to the late L. McCarthy (to whose memory this article is dedicated) for encouraging him to pursue this analysis. Thanks also are due L. Friedman and T. Takeuchi, for comments regarding an earlier version of this work.